\documentclass[twoside,twocolumn,9pt]{article}
\usepackage{extsizes}
\usepackage[super,sort&compress,comma]{natbib} 
\usepackage[version=3]{mhchem}
\usepackage[left=1.5cm, right=1.5cm, top=1.785cm, bottom=2.0cm]{geometry}
\usepackage{balance}
\usepackage{mathptmx}
\usepackage{sectsty}
\usepackage{graphicx} 
\usepackage{lastpage}
\usepackage[format=plain,justification=justified,singlelinecheck=false,font={stretch=1.125,small,sf},labelfont=bf,labelsep=space]{caption}
\usepackage{float}
\usepackage{fancyhdr}
\usepackage{fnpos}
\usepackage[english]{babel}
\addto{\captionsenglish}{%
  
}
\usepackage{array}
\usepackage{droidsans}
\usepackage{charter}
\usepackage[T1]{fontenc}
\usepackage[usenames,dvipsnames]{xcolor}
\usepackage{setspace}
\usepackage[compact]{titlesec}
\usepackage{hyperref}

\usepackage{epstopdf}

\definecolor{cream}{RGB}{222,217,201}

\begin{document}

\pagestyle{fancy}
\thispagestyle{plain}
\fancypagestyle{plain}{
\renewcommand{\headrulewidth}{0pt}
}

\makeFNbottom
\makeatletter
\renewcommand\LARGE{\@setfontsize\LARGE{15pt}{17}}
\renewcommand\Large{\@setfontsize\Large{12pt}{14}}
\renewcommand\large{\@setfontsize\large{10pt}{12}}
\renewcommand\footnotesize{\@setfontsize\footnotesize{7pt}{10}}
\makeatother

\renewcommand{\thefootnote}{\fnsymbol{footnote}}
\renewcommand\footnoterule{\vspace*{1pt}%
\color{cream}\hrule width 3.5in height 0.4pt \color{black}\vspace*{5pt}} 
\setcounter{secnumdepth}{5}

\makeatletter 
\renewcommand\@biblabel[1]{#1}            
\renewcommand\@makefntext[1]%
{\noindent\makebox[0pt][r]{\@thefnmark\,}#1}
\makeatother 
\renewcommand{\figurename}{\small{Fig.}~}
\sectionfont{\sffamily\Large}
\subsectionfont{\normalsize}
\subsubsectionfont{\bf}
\setstretch{1.125} 
\setlength{\skip\footins}{0.8cm}
\setlength{\footnotesep}{0.25cm}
\setlength{\jot}{10pt}
\titlespacing*{\section}{0pt}{4pt}{4pt}
\titlespacing*{\subsection}{0pt}{15pt}{1pt}

\fancyfoot{}
\fancyfoot[LO,RE]{\vspace{-7.1pt}\includegraphics[height=9pt]{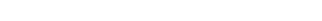}}
\fancyfoot[CO]{\vspace{-7.1pt}\hspace{13.2cm}\includegraphics{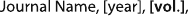}}
\fancyfoot[CE]{\vspace{-7.2pt}\hspace{-14.2cm}\includegraphics{head_foot/RF}}
\fancyfoot[RO]{\footnotesize{\sffamily{1--\pageref{LastPage} ~\textbar  \hspace{2pt}\thepage}}}
\fancyfoot[LE]{\footnotesize{\sffamily{\thepage~\textbar\hspace{3.45cm} 1--\pageref{LastPage}}}}
\fancyhead{}
\renewcommand{\headrulewidth}{0pt} 
\renewcommand{\footrulewidth}{0pt}
\setlength{\arrayrulewidth}{1pt}
\setlength{\columnsep}{6.5mm}
\setlength\bibsep{1pt}

\makeatletter 
\newlength{\figrulesep} 
\setlength{\figrulesep}{0.5\textfloatsep} 

\newcommand{\topfigrule}{\vspace*{-1pt}%
\noindent{\color{cream}\rule[-\figrulesep]{\columnwidth}{1.5pt}} }

\newcommand{\botfigrule}{\vspace*{-2pt}%
\noindent{\color{cream}\rule[\figrulesep]{\columnwidth}{1.5pt}} }

\newcommand{\dblfigrule}{\vspace*{-1pt}%
\noindent{\color{cream}\rule[-\figrulesep]{\textwidth}{1.5pt}} }

\makeatother

\twocolumn[
  \begin{@twocolumnfalse}
{\includegraphics[height=30pt]{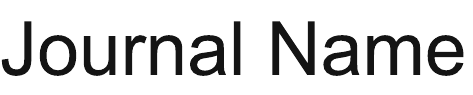}\hfill\raisebox{0pt}[0pt][0pt]{\includegraphics[height=55pt]{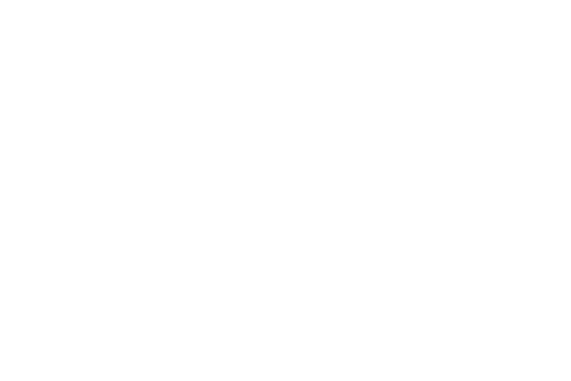}}\\[1ex]
\includegraphics[width=18.5cm]{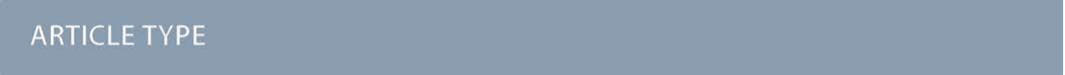}}\par
\vspace{1em}
\sffamily
\begin{tabular}{m{4.5cm} p{13.5cm} }

\includegraphics{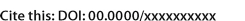} & \noindent\LARGE{\textbf{Origami Nano-gap Electrodes for Reversible Nanoparticle Trapping$^\dag$}} \\
\vspace{0.3cm} & \vspace{0.3cm} \\

& \noindent\large{Itir Bakis Dogru-Yuksel$^{\ast}$\textit{$^{a}$}, Allard P. Mosk\textit{$^{a}$}, and Sanli Faez\textit{$^{a}$}} \\

\includegraphics{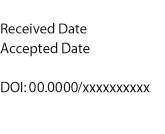} & \noindent\normalsize{We present a facile desktop fabrication method for origami-based nano-gap indium tin oxide (ITO) electrokinetic particle traps, providing a simplified approach compared to traditional lithographic techniques and effectively trapping of nanoparticles. 
Our approach involves bending ITO thin films on optically transparent polyethylene terephthalate (PET), creating an array of parallel nano-gaps. 
By strategically introducing weak points through cut-sharp edges, we successfully controlled the spread of nano-cracks. 
A single crack spanning the constriction width and splitting the conductive layers forms a nano-gap that can effectively trap small nanoparticles after applying an alternating electric potential across the nanogap. 
We analyze the conditions for reversible trapping and optimal performance of the nano-gap ITO electrodes with optical microscopy and electrokinetic impedance spectroscopy.
Our findings highlight the potential of this facile fabrication method for the use of ITO at active electro-actuated traps in microfluidic systems.} 

\end{tabular}

\end{@twocolumnfalse} \vspace{0.6cm}
]

\renewcommand*\rmdefault{bch}\normalfont\upshape
\rmfamily
\section*{}
\vspace{-1cm}


\footnotetext{\textit{$^{a}$~Nanophotonics, Debye Institute for Nanomaterials Science, Utrecht University, 3584 CC Utrecht, The Netherlands; E-mail: i.b.dogruyuksel@uu.nl, s.faez@uu.nl}}

\footnotetext{\dag~Electronic Supplementary Information (ESI) available: [details of any supplementary information available should be included here]. See DOI: 10.1039/cXsm00000x/}





Potentiodynamic manipulation and steering of nanoparticles in microfluidic systems is one of the key design elements for lab-on-chip applications.
Using alternating electrical potential for nanoparticle trapping is advantageous for higher electrochemical stability of the electrodes. The main mechanisms that have been used to achieve stable and reversible trapping of nanoparticles are dielectrophoresis (DEP) and alternating current electroosmotic (ACEO) flow ~\cite{wang2013mapping}, which may coexist in some geometries.
Dielectrophoresic forces arise from the interaction between the induced dipole moment of dielectric particles and the local gradient of the electric field in the surrounding environment~\cite{Lu2020}.
This interaction results in a net force that can be used to attract and trap particles within a non-uniform electric field near an electrode~\cite{Sun2019, Zaman2021}. 
For nanoscale localization and trapping of nanoparticles with DEP, e.g. to overcome the Brownian forces, the gradient must be significant on the scale of nanoparticle and that of the desired trapping area~\cite{pandey2018probing}. 

ACEO trapping is more complex than DEP as it depends on the 3-d flow created on top of the electrodes are around edged or at the inter-junction between conducting electrodes and charged dielectric surfaces.
This phenomenon often creates some net vorticity in the fluid flow that can draw the particles towards the junction, or can used for fluid mixing and creating an electroosmotic pump~\cite{Squires2009, Zhang_softmatter_2020}.
Both DEP and ACEO phenomena have been used for creating potentiodynamic nanoparticle traps and are widely used in lab-on-chip systems for the precise measurement~\cite{Pakhira2022, liu2015particle}, separation~\cite{Hughes2002,Liu2011, zhou2005lateral}, sorting~\cite{Boettcher2006, islam2006microfluidic}, and manipulation~\cite{Miled2012, Dalili2020, hilber2008particle} of suspended particles in liquid media. 
Because of electrical control and possibility of system integration, potentiodynamic manipulation techniques also enables accurate single-cell manipulation, contributing to advancements in biomedical diagnostics~\cite{Lambert2021,Hochstetter2020,Chuang2015} and therapy~\cite{Caglayan2020,ivanoff2013enhanced} featuring promising prospects for the development of a point-of-care tool in the future~\cite{demircan2013dielectrophoresis}.
For both mechanism, the key to realization of a sufficiently strong trap for small nanoparticles is to create a large gradient, hence a small radius of curvature at the electrode and small electrode separations are crucial\cite{chiou2005massively} and the effective force is inversely proportional to the gap size for micro-structured planar electrodes~\cite{Vezenov2011}.
Creating ACEO and DEP traps that are compatible with optical microscopy can put further constraints the choice of materials and requires a meticulous design of the nanoelectrodes.
Crafting robust electrodes with nanoscale edges demands precise engineering, and is often done with nanolithographic techniques to achieve trapping of for example sub 30~nm (bio)particles. 
Barik et al. employed atomic layer lithography to generate nanoscale gaps~\cite{barik2016ultralow}. 
Han et al. utilized electron beam lithography for fabricating an electrode array~\cite{han2022scalable}. 
Yu et al. employed photolithography to construct vertical nanogap architectures, showcasing precise nanoparticle capture and spatiotemporal manipulation~\cite{yu2020precise}.
While nanolithography provides a systemic path to optimization of such potentiodynamic traps, it is not accessible to all labs and requires specialised personnel and equipment. 

\begin{figure}[h] 
\centering
 \includegraphics[width=8cm]{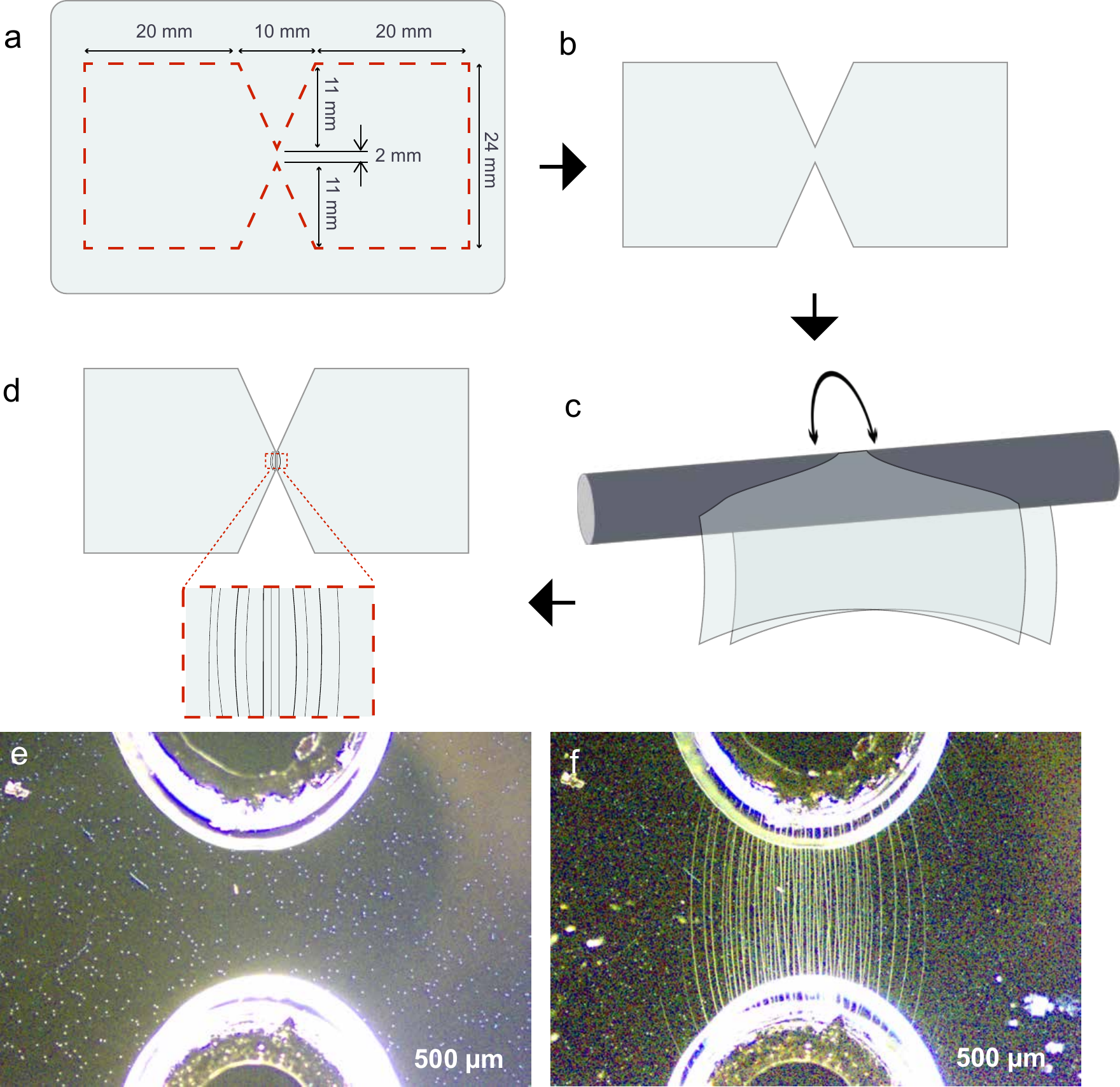}
  \caption{Fabrication of nano-gap origami electrodes using ITO on PET (a) The laser-cut design, with dashed red lines indicating the cut sections. (b) The cut layer after the laser-cutting process. (c) The process of fabricating origami nanocracks by bending the material around a metal rod. (d) The resulting nanocracks formed in the mid-region. The optical microscopy image of nanocrack fabrication region (e) before and (f) after bending.}
  \label{fgr:electrodes}
\end{figure}

In this article, we demonstrate an innovative method for generating nanoparticle-trapping nanogap electrodes that remarkably requires only regular office equipment. 
Our traps use the origami technique, involving the straightforward manipulation of thin conductive layers through bending. 
Our nanogap electrode production protocol stands out for its simplicity, speed, and reproducibility and can enable a wide community of researchers to use the advantages of electro-actuated trapping for their lab-on-chip applications.  

This fabrication method is inspired by previous work on origami-fabrication of quasi-ordered nano-cracks within protein layers, proposed as a mechanism for building distributed feedback bio-compatible laser~\cite{Dogru2021}. 
We apply this fabrication method to a thin indium tin oxide (ITO) coated on PET (Polyethylene Terephthalate). 
This substrate is widely available and affordable because of its commercial application for liquid crystal displays and devices. 
By implementing controlled crack propagation, a phenomenon extensively investigated before~\cite{Park2016,Yung2019,Park2019,Suh2015,Jung2020,Kim2016,Won2020, dogru2020high}, we demonstrate reversible trapping and particle alignment dynamics. 
The transparency of the substrate holds the potential to facilitate detailed examinations of biological structures, down to the level of individual cells or sub-cellular structures.

\section{Results}
\subsection{Nano-gap ITO electrodes}
We use a combination of laser-cutting and controlled bending to fabricate our origami nanogap electrodes. 
The substrate used for this article is a 90-150 Å ITO layer deposited on 200~$\mu$m PET with a resistance range of 350 - 500 $\Omega/\sqrt{\mathrm{cm}}$. 
The substrate flexibility is essential for the fabrication of origami-based electrodes, as it can be easily bent and manipulated to achieve the desired structure. 
We first cut a butterfly-shaped piece from the ITO-PET film. 
On purpose, we introduced weak points at the mid-region, where small triangles point towards the center with a 2~mm spacing as illustrated in Figure~\ref{fgr:electrodes}-a and as depicted in Figure~\ref{fgr:electrodes}-b.
This design reduces the likelihood of crack propagation further from the thin bridge. 
The resulting cracks align parallel to each other, which increases the probability of their completeness without overlapping with adjacent gaps. 
While laser cutting is handy for reproducibility, we have successfully replicated these experiments using scissor-cut samples. 
For the purpose of reproducibility, we avoid using scissor cutting, which might create additional cracks at the cut edges.
Remarkably, the presence of a single complete crack that spans the whole width of the narrow constriction and effectively dividing the conductive layer into at least two planar sections, is sufficient to form the nanogap trap for nanoparticles. 
To apply a uniform stress, we use a metal rod with a diameter of 60 mm to gently bend the PET film and roll it (see Figure~\ref{fgr:electrodes}-c). 
After this step, nanocracks are visible under an optical microscope.
They are formed in parallel to each other as illustrated in Figure~\ref{fgr:electrodes}-d, Figure~\ref{fgr:electrodes}-e and Figure~\ref{fgr:electrodes}-f present the state before and after bending, respectively. 

The gap size can be estimated by a simple mode that considers the difference in circumferences of the tangent circles on top and bottom of the the rounded film. 
The separation between adjacent cracks is relatively constant in the investigation area at $w_r\approx 15.5 \pm 4.4 \mu$m (Figure S1).
The nano-gap size is given by $d_g = t_f w_r/R$, with $R$ the bending radius and $t_f$ is the PET-substrate film thickness. Using a $R = 30$~mm and $t_f = 200 ~\mu$m, we estimate an average gap opening of $d_g = 100 \pm 30$~nm, which is slightly smaller than the gap size measured with electron microscopy. 
Our simple model considers only geometric effect and does not include the possible influence of possible shear tension between the ITO film and the PET substrate, or the irreversible deformation of the PET substrate after bending.
This simple model, however, shows how to control the gap size of the origami nano-crack electrodes by controlled bending of the substrate. 

\subsection{Reversible potentiodynamic trapping}

\begin{figure*}[h] 
\centering
  \includegraphics[width=15cm]{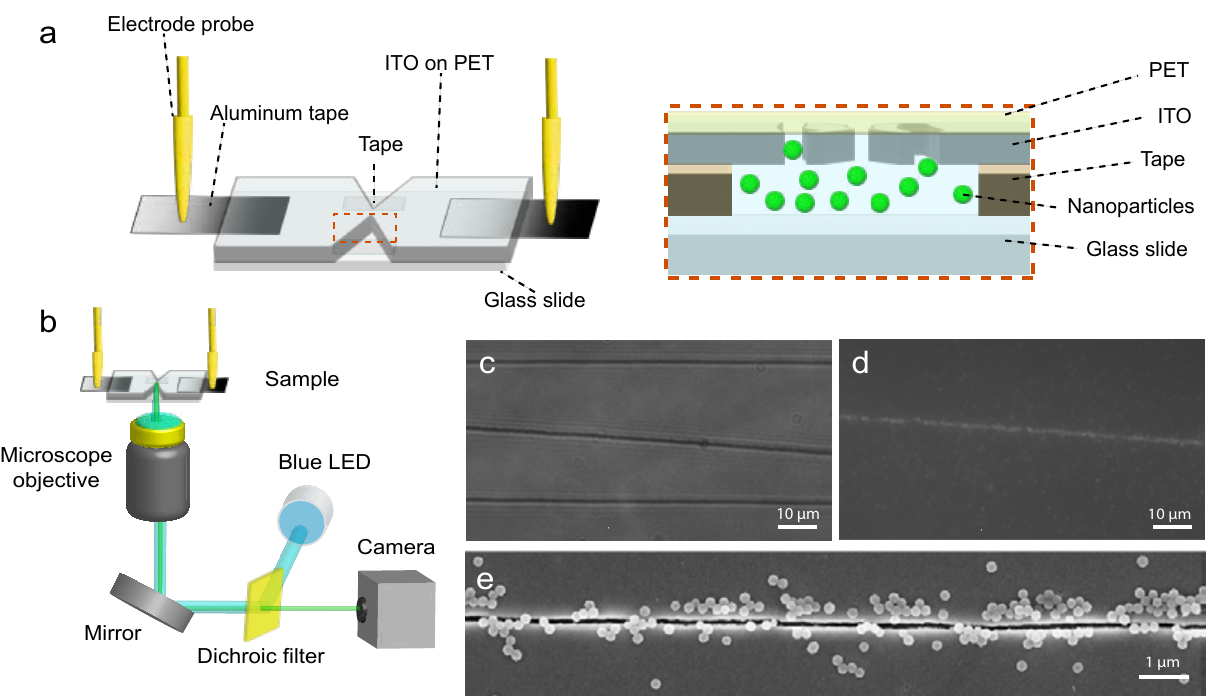}
  \caption{(a) Schematic representation of an inverted nanocrack electrode configuration, where nanoparticles in water are confined between the electrode and a glass slide, and an electric field is generated by aluminum tape and gold electrodes. (b) Schematic of a custom-built fluorescence microscopy setup and wavefunction generator connected to the gold electrodes. (c) Visualizing origami nanocracks under white light and (d) detecting particles trapped within a nanocrack using fluorescence. (e) SEM image depicting particles concentrated at the nanocrack edges.}
  \label{fgr:trapping}
\end{figure*}
The origami nanogap electrode was affixed to a glass slide using double-sided tape.
Electric contacts were made with aluminum tape to facilitate attachment of probing electrode rods from both sides (Figure~\ref{fgr:trapping}-a).
We emphasize that our samples, because of the transparent substrate, are compatible with any upright or inverted microscope. 
We used 200~nm fluorescent polystyrene (PS) nanoparticles in water to demonstrate reversible trapping around the gaps. 
To visualize the particles, we used a lab wave-function generator to apply the trapping potential and a custom-built fluorescence microscopy setup, with a blue LED for illumination, to observe their response (Figure~\ref{fgr:trapping}-b).
Effective trapping occurs when the electrokinetic forces surpass counteracting forces like thermal Brownian motion or convection~\cite{Ghomian2022}.
When cracks are shorter than the cross section, the negligible voltage drop due to incomplete electric conduction has no observable influence on the particle motion. 
The Scanning Electron Microscopy (SEM) images in Figure S2 provides additional images confirming the completeness of the trapping origami nanocrack. 
This effect is highlighted in the bright-field image of three cracks in Figure~\ref{fgr:trapping}-c. Moreover, the fluorescent-micrsocopy image in Figure~\ref{fgr:trapping}-d from the same region as Figure~\ref{fgr:trapping}-c clearly shows that incomplete cracks do not achieve the desired effect.The SEM image in Figure~\ref{fgr:trapping}-e demonstrates adhesion of particles on the edges of the crack that cover the entire cross-section of the constriction under specific potential parameters.

\begin{figure*}[h] 
\centering
  \includegraphics[width=15cm]{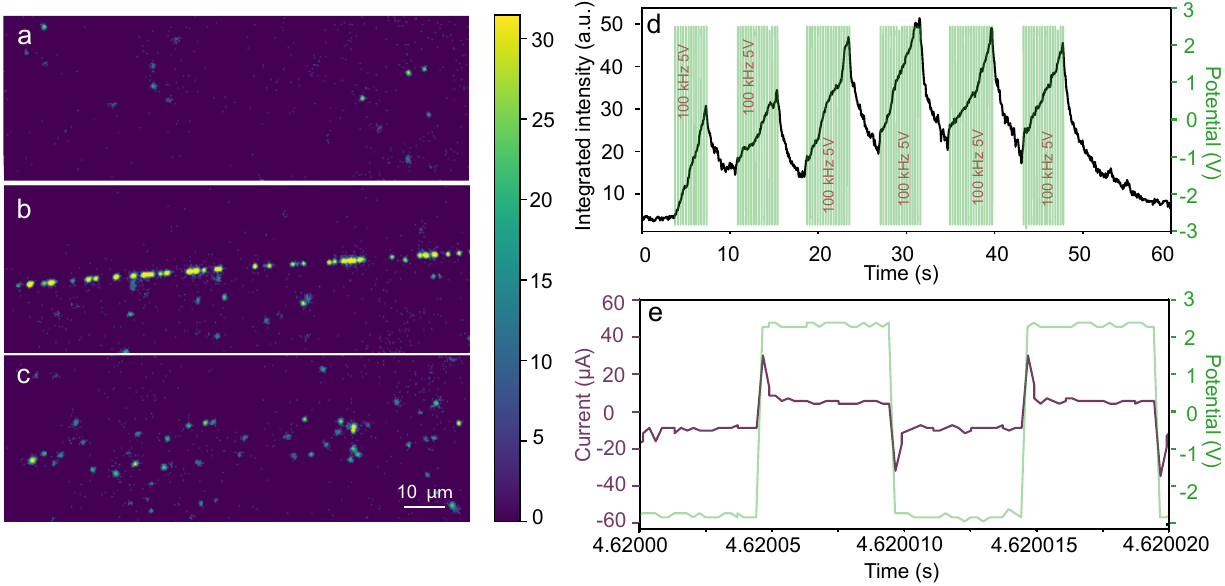}
  \caption{Dynamic stages of particles in water within nanocracks: (a) before, (b) during, and (c) after the application of an electric field. (d) Integrated intensity change over time (black) is correlated with the applied potential (green). (e) A close-up view of the applied potential (green) with 100-microsecond resolution, alongside simultaneous recording of the electric current passing through the nanocrack samples (purple).  }
  \label{fgr:action}
\end{figure*}

Before applying the electric field, we observe the particles exhibiting Brownian motion, as depicted in Figure~\ref{fgr:action}-a. When the electric field is first applied (a 5~V peak-to-peak rectangular waveform at 100~kHz), the particles are drawn to and remain localized at the nanocrack region, as illustrated in Figure~\ref{fgr:action}-b. However, upon deactivation of the field, the particles are promptly released, as shown in Figure~\ref{fgr:action}-c (Also see Supplementary Video 1). 
The rapid response of particles to an applied electric field is assessed by measuring the integrated intensity change along a nanocrack.
The integrated intensity undergoes a rapid increase upon applying the electric field application, followed by a relatively slower decline when the potential is set back to zero, as depicted in Figure~\ref{fgr:action}-d. 
The green data represent the applied potential, demonstrating that the AC electric potential is intermittently applied for 5-second time-interval, repeated six times with 3-second intervals of zero potential. 
The zoomed-in view in Figure~\ref{fgr:action}-e, focusing on an applied potential of 5V at 100 kHz with a temporal resolution of 100 microseconds, enables a detailed analysis of the electric field dynamics while the purple data show the current measured from the nanocracks. 

\begin{figure}[h] 
\centering
  \includegraphics[width=8cm]{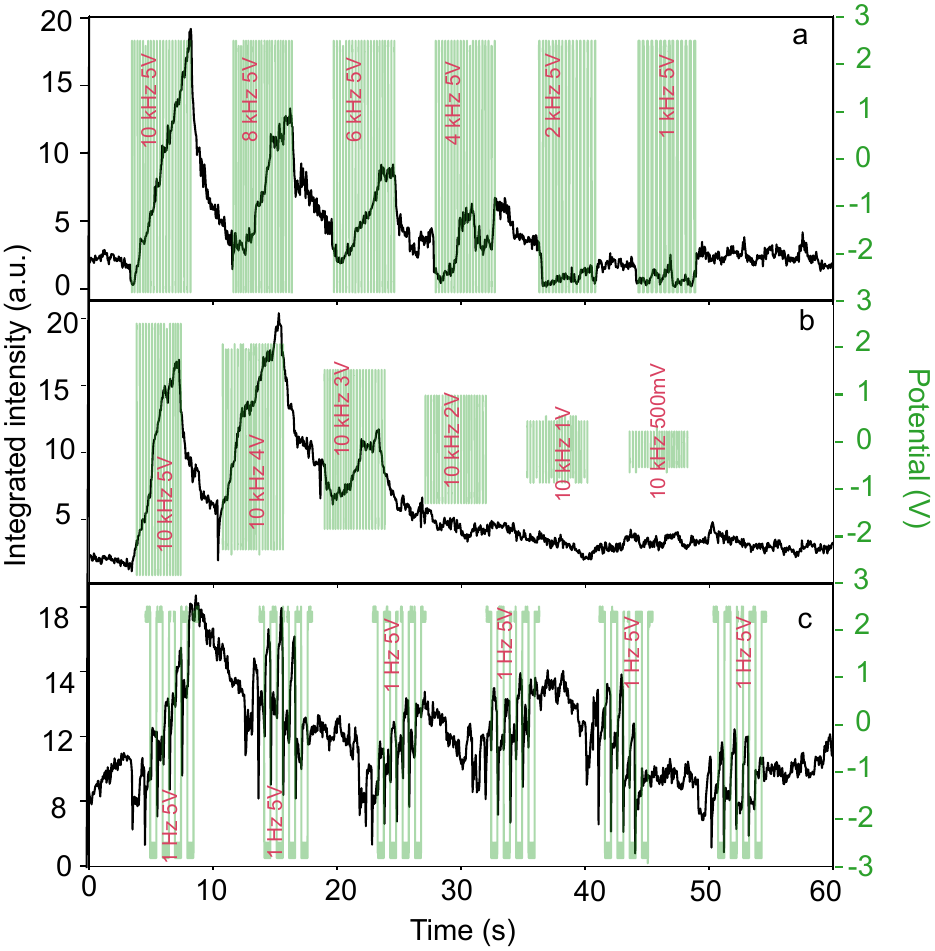}
  \caption{The integrated intensity change over time (a) as the frequency is systematically decreased, (b) when the frequency is decreased (b) when the applied voltage is systematically decreased and (c) at 5V with 1 Hz application. }
  \label{fgr:freq}
\end{figure}

Under a 5-V peak-to-peak amplitude of the applied potential we gradually reduced the frequency from 100~kHz to 1~Hz (See also Supplementary Video 2).It is noteworthy to mention that when two complete cracks are present, we observed simultaneous trapping occurring within these parallel cracks as shown in Figure S3.The nuanced dynamics of reversible trapping at 10,000 Hz are further elucidated in Figure S4, resembling Figure~\ref{fgr:action} but with a diminished integrated intensity. 
Nanoparticle trapping persisted at all frequencies above 2~kHz. 
We observed no trapping at frequencies lower than 2~kHz (Figure~\ref{fgr:freq}-a). 
The specifics of particle behavior spanning from 100 kHz to 1 Hz (100 kHz, 10 kHz, 1 kHz, 100 Hz, 10 Hz, and 1 Hz) are presented in Figure S5.

Conversely, when holding the frequency at 10~kHz and varying the amplitude (5, 4, 3, 2, 1 V, and 500 mV), a distinct pattern emerged~(Figure~\ref{fgr:freq}-b). 
Trapping effects were negligible at or below 2~V, while reversibiler trapping is observed at higher voltages (see Suppementray Video 3). 

Intriguingly, under the application of 5V at 1 Hz (cycled six times over 5 seconds with 3-second intervals), particles exhibited a unique response. 
They displayed a oscillatory motion, shuttling between sides of the nanocrack line in sync with the applied potential. 
This dynamic movement is clearly depicted in Figure~\ref{fgr:freq}-c, aligning with the peak and dip points corresponding to the application of the electric field, which is a clear indication of the bulk-flow induced by applying the electric potential. 
However, the cycle-averaged forces at these low frequency are not enough to concentrate or eventually trap the nanoparticles (see also Supplementary Video 4).

The observed electrokinetic trapping cannot be explained by DEP as we expect positive (repulsive from the electrode) forces on polystyrene particles at low frequencies. 
The DEP forces are expected to change sign at frequencies above 100~kHz for the conditions of our sample~\cite{wang2013mapping}.
However, we can observe trapping at much lower frequencies, even down to 2~kHz.
However, ACEO trapping has been observed at lower frequencies and we anticipate is the underlying mechanism for trapping the nanoparticles in our system. 
To explore this trapping conditions and stability of the sample empirically, we conducted a parametric study of applied potential and frequencies using cyclic voltammetry and electrochemical impedance spectroscopy (EIS) of the same nanogap electrodes in contact with the nanoparticle sample or other electrolyte solutions.

\subsection{Impedance spectroscopy of origami nanocrack electrodes}
In order to characterize and understand the ideal trapping conditions, and check the reproducibility of each fabricated electrode, we use electrochemical impedance spectroscopy (EIS).
In this method, the complex-valued conductivity of the sample is measured as a function of frequency and amplitude for a sinusoidal AC potential, revealing the various conduction regimes in the electrolyte. 
To understand the influence of the electric-double-layer formation at the electrodes, we also repeat our characterization for a series of ionic strength of in solutions of increasing potassium chloride (KCl) concentration.

\begin{figure}[h] 
\centering
   \includegraphics[width=8cm]{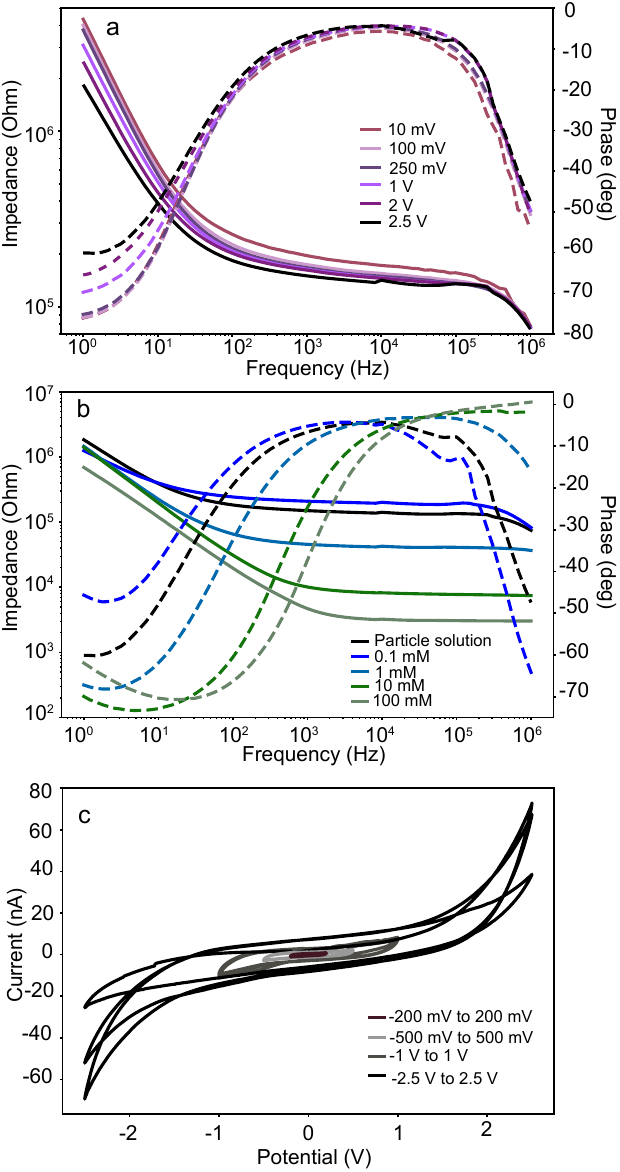}
  \caption{Bode plots of the impedance of the ITO electrodes for sinusoidal alternating potential (a) in contact with the nanoparticle solution for varying wave amplitude, (b) in contact with a KCl solution of different concentration. (c) Cyclic voltammetry of ITO electrodes in contact with the nanoparticle solution for different ranges of the applied potential. For each range 3 cycles have been measured. Only the highest range displays some alterations of the measured current for the extremes of the potential that can be attributed to electrochemical change of the surface properties such as roughness.}
  \label{fgr:imped}
\end{figure}

Fig.~\ref{fgr:imped}(a) depicts the measured impedance spectrum of the ITO electrodes in contact with the nanoparticle suspension for various applied potentials.
We can observe little difference alteration in the impedance response, especially at frequencies above 100~Hz, which testify to the electrochemical stability of the ITO substrate.

In Fig.~\ref{fgr:imped}(b) we present the EIS measurements for the ITO electrodes for various concentrations of KCl dissolved in water at $2.5$~V. 
We can observe that the conductivity decreases (impedance magnitude increases) for higher salt concentration.
Furthermore, the response phase at the highest frequencies deviates from zero for the lowest measured salt concentration of 0.1 mM, for which the charging time of the electric double layer becomes comparable with the period of the waveform.
The response of the nanoparticle solution is closest to that of the 0.1~mM salt concentration.
We therefore conclude that the residual ionic strength for the nanoparticle solution used for our measurements is in that range.
The cyclic voltammery diagram of origami electrodes with particle solution depicted in Fig.~\ref{fgr:imped}(c) also confirm the electrochemical stability for the potential range of $\pm 2.5$~V. 
Water splitting reactions speed up outside this range, resulting in higher Faradaic currents, but this effect seems to be negligible for AC potentials at frequencies larger than 100~Hz that are suitable for trapping at the nano-gap and the main focus of this article.
The absence of Faradaic reactions is advantageous for the main function of the nano-gap trapping as it enhances the chemical stability of the ITO electrode. 
The stability conditions, however, are only tested at neutral pH. 
Use of acidic or basic solutions might further limit the usability condition of this device.

\section{Conclusions}

In this study, we introduce a novel approach to nanogap electrode fabrication, utilizing origami principles to create controlled nanocracks in transparent and conductive films. This method deviates from conventional lithographic techniques, offering point-of-use one-step process for the production of parallel-arranged nanogap electrode patterns.
The stress-induced merging of strategically introduced weak points results in reversible nanoparticle confinement, as evidenced by the observed dynamics under varying parameters such as amplitude and frequency of the applied waveform.
The trapping behavior persists at AC frequencies as low as 2~kHz, consistent with the hypothesis of AC electroosmosis as the main underlying mechanism.
Based on EIS measurements, we demonstrate the stability of ITO electrodes for the relevant potentials and frequencies and rule out the influence of Faradaic reactions at neutral pH, which could otherwise result in chemical degradation.

These nanogap electrodes can be used in various applications, including integration into lab-on-a-chip systems for controlled analysis of biological particles, microfluidic devices for efficient particle transport and sorting, and biotechnology applications for the manipulation of cells and biomolecules.

\section*{Materials and Methods}
\subsection{Nano-gap ITO electrode fabrication}
90-150Å ITO coated PET films with a thickness of 200 µm (OCF2520, Thorlabs) were precision-cut using a GCC LaserPro X380 Laser Engraver/Cutter according to a vectorial program defining the electrode design. The resulting electrodes measured 24 mm in height and 50 mm in length, featuring weak points at the mid-region with small triangles pointing towards the center, spaced at 2 mm intervals in a ribbon shape. Nanocracks were induced in the ITO-coated PET film by bending it around metal rods with a constant 30 mm radius of curvature. The longitudinal strain applied to the film surface during smooth rolling resulted in uniformly spaced nanocracks with an average of 15.5 µm intervals in the mid region. The electrodes were inverted and affixed to a 24x50 mm No. 1.5 glass slide using small square pieces of double-sided tape. The ITO conductivity was enhanced with aluminum tape, facilitating the attachment of electrode probes from both sides. This assembly effectively confined the particle solution between the glass slide and the electrode design, minimizing the impact of water evaporation and drift. 

\subsection{Particle solution preparation}
Aqueous suspensions of 0.20 µm Fluoresbrite® YG Carboxylate Microspheres (2.5\% w/v) with a maximum excitation of 441 nm and maximum emission of 486 nm were diluted 10,000 times with Milli-Q water. For trapping experiments, 60 µl of the diluted solution was used. 

\subsection{Trapping Experiments}
For electrokinetic trapping measurements we used a function generator and a home-built fluorescence microscopy setup. A blue LED (Thorlabs, M395L4) was employed to excite the particles, with the excitation light reflected by a dichroic filter directed to the microscope objective (ZEISS 40X/0.65) via a mirror. The emitted light followed the reverse path, passing through a long-pass dichroic filter (500 nm) and reaching the camera (Hamamatsu Digital Camera, C11440). This configuration effectively filtered the source light from the image, enabling particle visualization. A waveform generator (KEYSIGHT InfiniiVision DSOX2024A) synchronized with the camera monitored the applied voltage and circuit, facilitating simultaneous observation of the impact of applied potential on particle behavior.
\subsection{Impedance Spectroscopy}
The BioLogic SP300 Potentiostat was employed for impedance and cyclic voltammetry measurements of the origami nanocrack electrodes.

\subsection{Characterization}
Nanocrack morphology and trapped PS particles were analyzed using a ZEISS EVO 15 SEM operated at 5 kV.
\section*{Author Contributions}
\textbf{Itir Bakis Dogru-Yuksel}: Visualization, Investigation, Conceptualization, Methodology, Writing- Original draft preparation.
\textbf{Allard P. Mosk}: Validation, Writing- Reviewing and Editing, Supervision.
\textbf{Sanli Faez}: Conceptualization, Methodology, Writing- Original draft preparation, Writing- Reviewing and Editing, Supervision.

\section*{Conflicts of interest}
``There are no conflicts to declare''.

\section*{Acknowledgements}
We thank Ivo Kootwijk and Sjoerd Kwaak for assistance in some of the earlier measurements for this project. We would like to acknowledge technical support with the measurement setup from Aron Opheij, Jan Bonne Aan, Dante Killian, Arjan Driessen and Paul Jurrius. This research was supported by Refeyn LTD and Nederlandse Organisatie voor Wetenschappelijk Onderzoek (Vici 68047618).



\balance


\bibliography{rsc} 
\bibliographystyle{rsc} 

\end{document}